\documentclass[aps,prl,twocolumn,groupedaddress,showpacs]{revtex4}
\usepackage{graphicx}
\usepackage{amsmath}
\usepackage{amssymb}
\usepackage{}

\begin{document}
\title{Multiphoton Rabi Oscillations of Correlated Electrons in Strong Field Nonsequential Double Ionization}
\author{Qing Liao, Yueming Zhou, Cheng Huang and Peixiang Lu$\footnote{Corresponding author: lupeixiang@mail.hust.edu.cn}$}
\affiliation{Wuhan National Laboratory for Optoelectronics and
School of Physics, Huazhong University of Science and Technology,
Wuhan 430074, P. R. China}
\date{\today}

\begin{abstract}
With quantum calculations, we have investigated the multiphoton
nonsequential double ionization of helium atoms in intense laser
fields at ultraviolet wavelengths. Very surprisingly, we find a
so-far unobserved double-circle structure in the correlated electron
momentum spectra. The double-circle structure essentially reveals
multiphoton Rabi oscillations of two electrons, which are strongly
supported by the oscillating population of a certain doubly excited
state and by the oscillating double ionization signals. This
two-electron multiphoton Rabi effect provides profound
understandings of electronic correlations and complicated
multiphoton phenomena and is expected to be a new tool for broad
applications, such as quantum coherent control.
\end{abstract}
\pacs{32.80.Rm, 32.80.Fb, 32.80.Wr} \maketitle Electronic
correlations are of fundamental importance to the dynamics of many
phenomena such as high-temperature superconductivity in solid
states. Nonsequential double ionization (NSDI) of atoms and
molecules by short intense laser pulses can provide one of the basic
examples for studies of dynamical electron correlations and thus has
been investigated extensively in both experiment \cite{Walker,Weber,
Feuerstein, Weckenbrock,Staudte} and theory \cite{Kopold, Lein,
Panfili, Faria} in the past few decades. Because of possessing rich
information, the correlated electron momentum spectra \cite{Weber,
Feuerstein, Weckenbrock,Staudte} have revealed a great many physical
pictures of electron-electron correlation in NSDI under recollision
mechanism. These physical pictures are well-described by the
classical recollision model \cite{Feuerstein,Corkum,Moshammer1}. The
correlated electron momentum spectra from quantum mechanical
calculations of NSDI of He at extreme ultraviolet (XUV) wavelengths
\cite{Parker}, as well as visible and ultraviolet (UV) wavelengths
\cite{Lein2,Parker2}, exhibit a circle (or circular arc) structure
with energy separation of the photon energy. This structure reveals
a resonant double ionization process in which the correlated
electrons simultaneously absorb and share energy in integer units of
the photon energy, transiting from the ground state into continuum
states \cite{Parker}. Even at near-infrared wavelengths there is
also a resonant double ionization process dominating NSDI of He
after doubly excited states populated via recollision below
recollision threshold \cite{Liao}. Such a NSDI process has been
observed in recent experiments on double ionization of He and Ne by
strong free-electron laser pulses at vacuum  UV wavelengths \cite
{Moshammer2, Rudenko2}.

Another fundamental effect in nonlinear light-matter interaction is
optical Rabi oscillations, which are of general importance to
quantum optics and have extensive applications in many fields such
as quantum coherent control in atomic clocks
\cite{Chaudhury,Windpassinger} and especially in quantum computing
\cite{Li,Collin,Wallraff}. In this Letter, we demonstrate
multiphoton Rabi effect of two strongly correlated electrons in NSDI
of He by strong laser fields at UV wavelengths. By numerically
solving the two-electron time-dependent Schr$\ddot{o}$dinger
equation, we obtained the correlated electron momentum spectra from
NSDI. A "one-plus-one"-dimensional model of a helium atom with soft
Coulomb interactions, where the motion of both electrons is
restricted to the laser polarization direction, is employed. This
model has been able to reproduce many NSDI features \cite{Lein2,
Lein, Lappas}. We use the split-operator spectral method \cite{Feit}
to numerically solve the two-electron time-dependent
Schr$\ddot{o}$dinger equation (in atomic units)
\begin{equation}\label{e1}
-i\frac{\partial}{\partial
t}\Psi(z_1,z_2,t)=H(z_1,z_2,t)\Psi(z_1,z_2,t),
\end{equation}
where $z_1, z_2$ are the electron coordinates. $H(z_1,z_2,t)$ is the
total Hamiltonian and reads
\begin{eqnarray}\label{e2}
H(z_1,z_2,t)&=&-\frac{1}{2}\frac{\partial^2}{\partial{z_1^2}}-\frac{1}{2}\frac{\partial^2}{\partial{z_2^2}}
-\frac{2}{\sqrt{z_1^2+1}}-\frac{2}{\sqrt{z_2^2+1}}\nonumber\\&&
+\frac{1}{\sqrt{(z_1-z_2)^2+1}}+(z_1+z_2)E(t).
\end{eqnarray}
$E(t)$ is the electric field of a laser pulse. Following Ref.
\cite{Lein}, the two-dimensional space is partitioned into two outer
regions: (A) $\{|z_1|<a\}$, or $\{|z_2|<a\}$ and (B)
$\{|z_1|,|z_2|\geq a\}$ with $a=150$ a.u. The final results are
insensitive to the choice of $a$ ranging from 100 to 200 a.u. In
region A, the wave function is propagated exactly in the presence of
combined Coulomb and laser field potentials. In region B, which
corresponds to double ionization, all the Coulomb potentials between
the particles are neglected and the time evolution of the wave
function can be performed simply by multiplications in momentum
space. The two regions are smoothly divided by a splitting technique
\cite{Tong}. At the end of the propagation, the wave function in
region B yields the two-electron momentum and energy spectra from
double ionization.

Our calculations use trapezoidally shaped laser pulses with a total
duration of 60 optical cycles, switched on and off linearly over 10
optical cycles respectively. A very large grid size of
$2500\times2500$ a.u. with a spatial step of 0.15 a.u. is used,
while the time step is 0.1 a.u. The very large grid provides
sufficiently dense continuum states \cite{Lein2} to yield highly
accurate two-electron momentum and energy spectra. The initiate wave
function is the two-electron ground state of He obtained by
imaginary-time propagation. After the end of the pulse, the wave
function is allowed to propagate without laser field for an
additional time of 40 optical cycles. The final results do not
change any more even though the wave function propagates without
laser field for a longer additional time.

\begin{figure}
\begin{center}
\includegraphics[width=7.5cm]{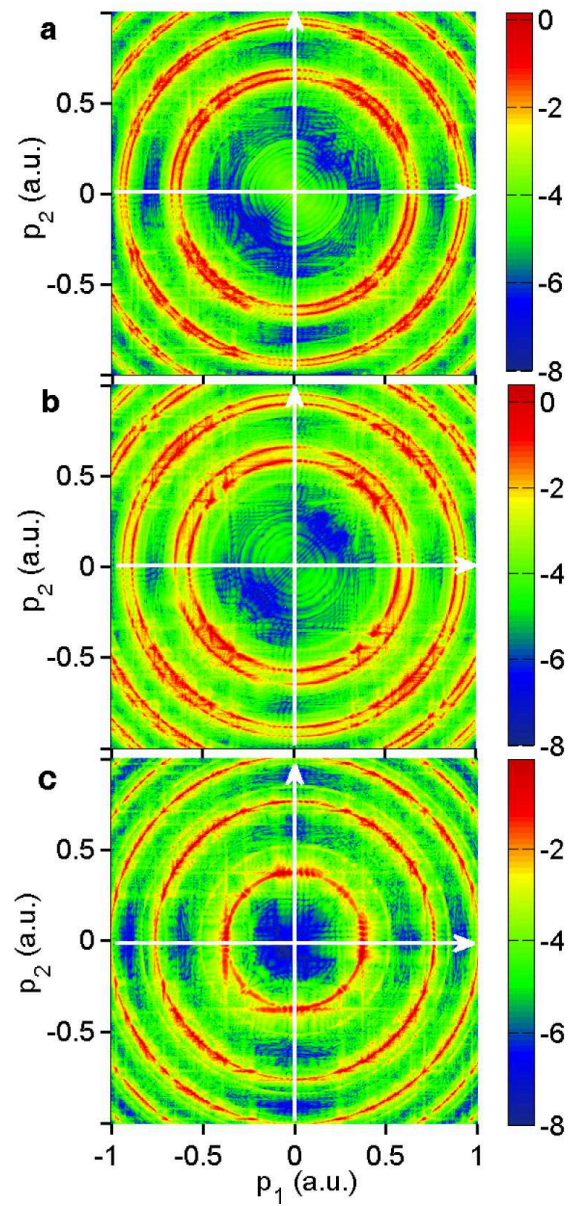}
\caption{\label{fig1}(color online) Log plot of the correlated
electron momentum spectrum for double ionization of He by laser
pulses at (a) 198 nm, 0.3 PW/cm$^2$, (b) 198 nm, 0.4 PW/cm$^2$ and
(c) 208 nm, 0.3 PW/cm$^2$. The units are arbitrary.}
\end{center}
\end{figure}

Figure 1 displays the resulting correlated electron momentum
spectrum from double ionization of helium atoms. Very surprisingly,
a double-circle structure is prominent in the momentum spectra for
double ionization at 198 nm, 0.3 PW/cm$^2$ [Fig. 1(a)] and 0.4
PW/cm$^2$ [Fig. 1(b)], which differs from the single-circle
structure in previous works \cite{Parker, Lein2, Parker2, Liao}.
These concentric circles satisfy $(p_1^2+p_2^2)=constant$, which is
the signature of a resonant double-ionization process
\cite{Parker2}. In this process, the two electrons simultaneously
absorb an integer number of photons and share the excess energy in
integer units of the photon energy. This process has been called
nonsequential double-electron above-threshold ionization
(DATI)\cite{Parker2}. Comparing Figs. 1(a) with 1(b), we find that
the separation of each doublet becomes larger when increasing the
laser intensity and keeping the wavelength unchanged. At 198 nm, 0.1
PW/cm$^2$, the separation becomes undistinguishable in the
correlated spectrum (not shown in this Letter). However, the
separation between each doublet is much less than that between
adjacent doublets. The relations between these circles manifest
themselves in the corresponding total kinetic energy spectra of two
ionized electrons, as shown in Fig. 2. The energy separation between
adjacent doublets is constant and equals the photon energy, whereas
the energy separation between each doublet is also constant for one
laser intensity but becomes larger with the increasing intensity. At
0.2 PW/cm$^2$, 0.3 PW/cm$^2$ and 0.4 PW/cm$^2$, they are, on
average, 0.013 a.u., 0.030 a.u. and 0.048 a.u., respectively.

\begin{figure}
\begin{center}
\includegraphics[width=8cm]{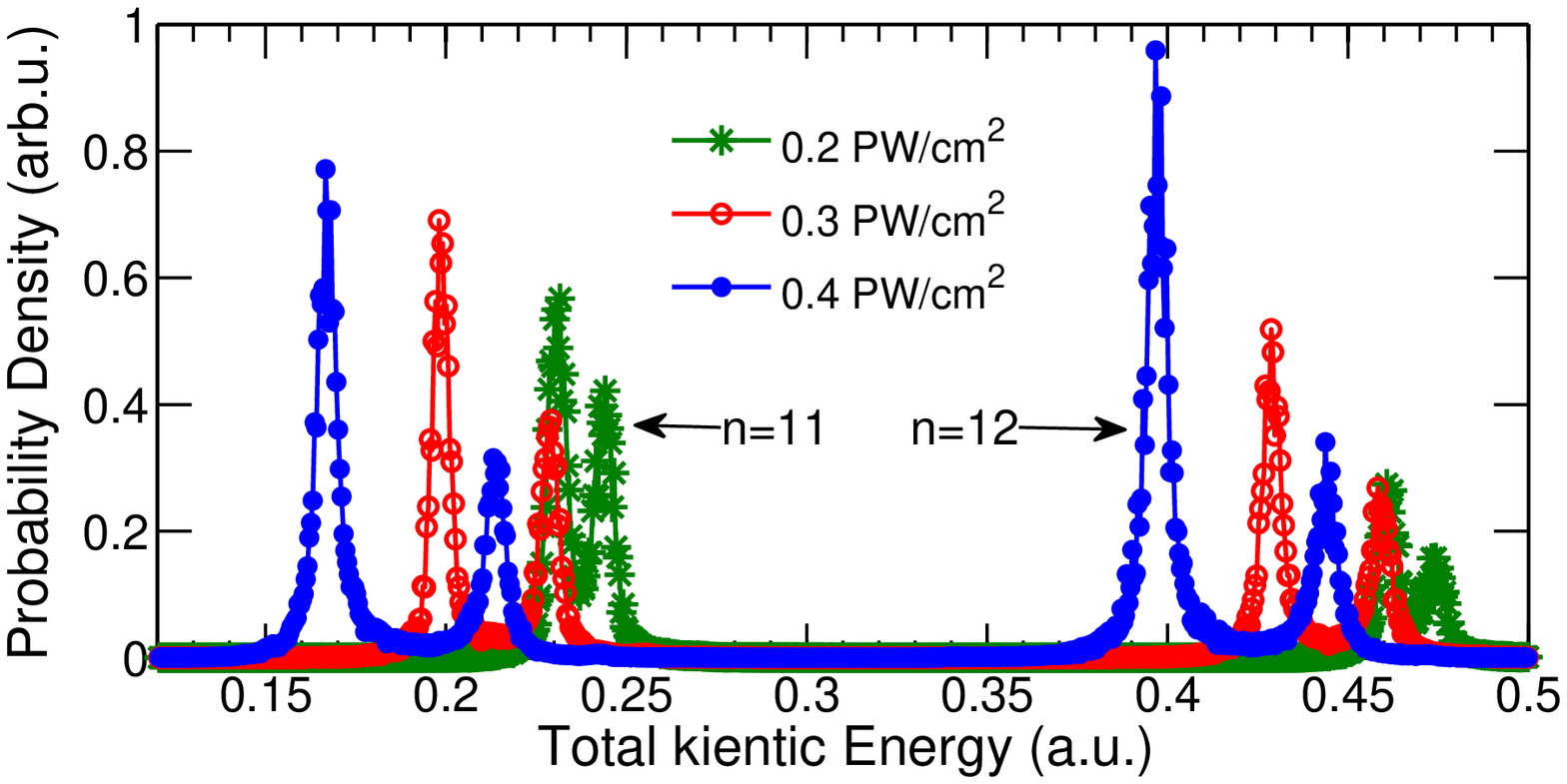}
\caption{\label{fig2}(color online) Photoelectron total-kinetic
energy spectrum of two ionized electrons from double ionization of
He by 198 nm laser pulses at different intensities.}
\end{center}
\end{figure}

In order to gain insight into the physical mechanism responsible for
the double-circle structure in the correlated momentum spectra, we
investigate the time evolution of the population of doubly excited
states and the flux of double ionization. The population of doubly
excited states is examined by monitoring the population in region 1:
$\{$7 a.u.$<|z_1|<$10 a.u., 7 a.u.$<|z_2|<$10 a.u. $\}$
\cite{Dundas} and we define region 2: $\{|z_1|>$20 a.u., $|z_2|>$20
a.u.$\}$ as the region of doubly ionizing wavepackets. We must
emphasize the fact that the doubly ionizing wavepackets may
contribute to the population in region 1. However, if the doubly
ionizing wavepackets arise dominantly from doubly excited states,
the contribution from doubly ionizing wavepackets to population in
region 1 can be neglected. This is verified for the case when the
double-circle structure dominates the correlated spectra, which is
elaborated below. Fig. 3 shows the population of doubly excited
states and the flux of double ionization as functions of time for
the laser parameters of 198 nm, 0.3 PW/cm$^2$ [Figs. 3(a) and 3(b)]
and 198 nm, 0.4 PW/cm$^2$ [Figs. 3(c) and 3(d)]. The population of
doubly excited states and the flux of double ionization oscillate
synchronously and damply. Note that the oscillating periods of the
population of doubly excited states and of the flux of double
ionization are equal. For higher laser intensities, the oscillating
period becomes shorter and the oscillation damping becomes larger.
The periods are, on average, 16.2, 7.64 and 4.92 optical cycles for
198 nm pulses with intensities of 0.2 PW/cm$^2$, 0.3 PW/cm$^2$ and
0.4 PW/cm$^2$, respectively. Thereby, the corresponding frequencies
are 0.014 a.u., 0.030 a.u. and 0.047 a.u., in very good agreement
with energy separations of the corresponding doublets.

Evidently, the above oscillations are the so-called Rabi
oscillations that occurs when there is resonance between the
two-electron ground state and a certain doubly excited state. Both
the two states are split into two quasi-energy states separated in
energy by the Rabi frequency \cite{Delone}. This is the physical
mechanism responsible for the double-circle structure in the
correlated momentum spectra. Because the two-electron ground state
population and the doubly excited state population is depleted by
single and double ionization, the Rabi oscillations are damped
strongly depending on the laser intensity. The Rabi frequency is
given by \cite{Loudon}
\begin{equation}\label{e3}
\Omega=\sqrt{(m\omega-\omega_0)^2+(\mu E_0)^2},
\end{equation}
where $\mu$ is the transition dipole moment, $E_0$ is the field
amplitude of the laser pulse, $\omega_0$ is the transition
frequency, $\omega$ is the photon frequency and $m$ is the number of
photon that resonance requires. With the assumption that the change
of $\omega_0$ is negligible relatively to that of $m\omega$ at the
condition of varying the laser wavelength slightly and keeping the
laser intensity constant, the value of $m$ can be determined. For
0.3 PW/cm$^2$ pulses, the Rabi frequency is 0.03 a.u., 0.038 a.u.
and 0.044 a.u. at 198 nm, 200 nm and 202 nm, respectively. According
to Eq. (\ref{e3}), we determine $m=6$.

\begin{figure}
\begin{center}
\includegraphics[width=8cm]{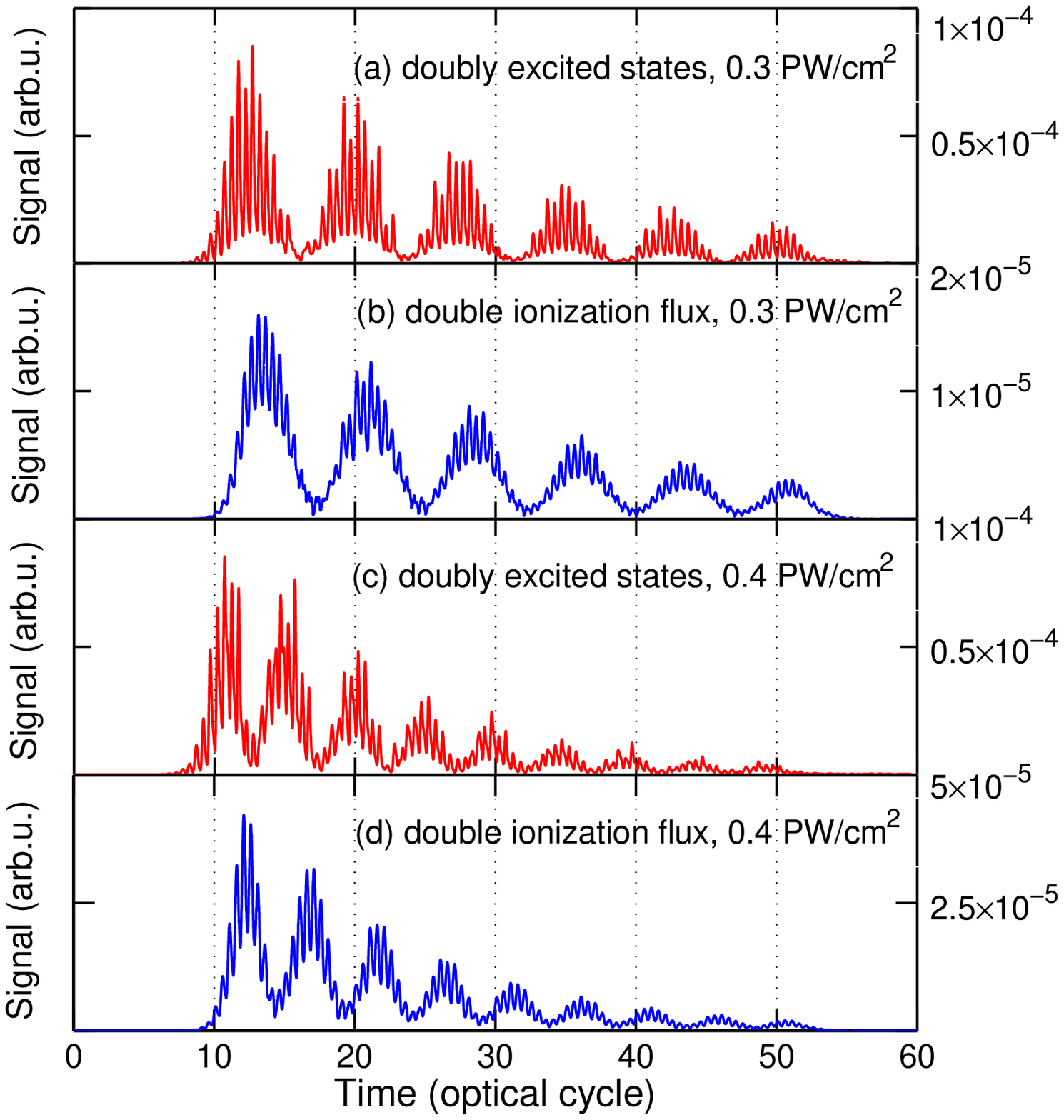}
\caption{\label{fig3}(color online) (a, c) Population of doubly
excited states and (b, d) double ionization fluxes as functions of
time. The laser parameters are 198 nm, 0.3 PW/cm$^2$ for (a, b) and
198 nm, 0.4 PW/cm$^2$ for (c, d), respectively.}
\end{center}
\end{figure}

\begin{figure}
\begin{center}
\includegraphics[width=8cm]{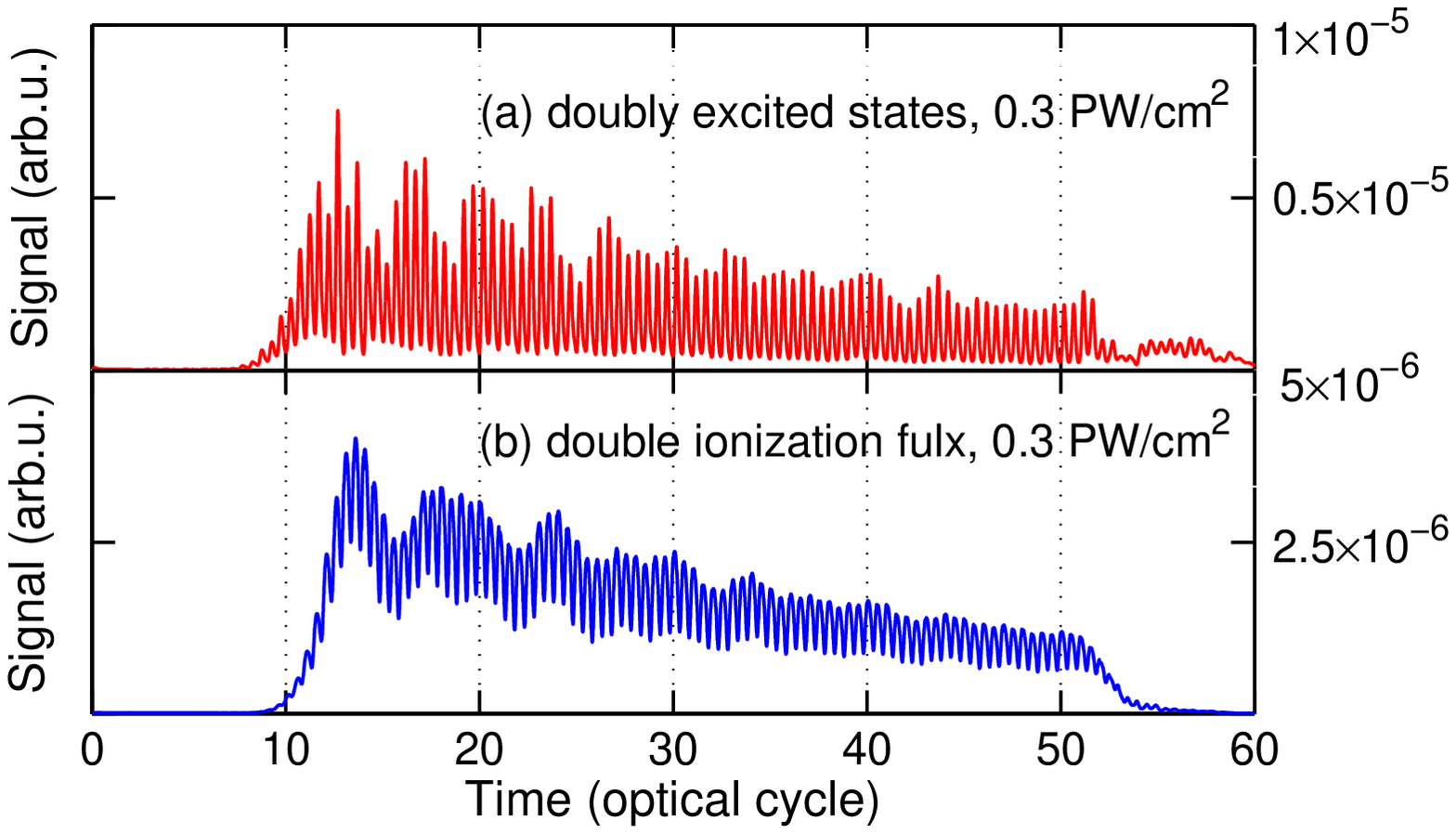}
\caption{\label{fig4}(color online)(a) Population of doubly excited
states and (b) double ionization fluxes as functions of time. The
laser parameters are 208 nm, 0.3 PW/cm$^2$.}
\end{center}
\end{figure}

Now it is very obvious that the double-cycle structure essentially
reveals two stages of the multiphoton double ionization process in
which the two electrons are strongly correlated. Firstly, both
electrons resonantly absorb a number of photons, transiting from the
two-electron ground state into a certain doubly excited state. Then
they emit photons, transiting into the ground state, or are emitted
by resonantly absorbing additional a number of photons and sharing
excess energy in integer units of the photon energy. In the whole
process, the two electrons behave in the same way as one electron.
Therefore, analogous to the kinetic energy of photoelectrons
resulting from above-threshold ionization by ultrashort pulses
\cite{Freeman}, the total kinetic energy of the doubly ionized
electrons can be approximately predicted by
\begin{equation}\label{e4}
E_n=n\hbar\omega-I_p-2U_p.
\end{equation}
$n$ is the total number of photon absorber by the two electrons.
$I_p=2.238$ a.u., is the double-ionization potential of He atom in
our model. The ponderomotive energy $U_p=E_0^2/(4\omega^2)$. The AC
Stark shift of the ground state is not included in Eq. (\ref{e4}).
Eq. (\ref{e4}) gives a very good estimation of the middle position
of the double-peak. For example, for n=11, Eq. (\ref{e4}) gives
0.242 a.u., 0.215 a.u. and 0.188 a.u. at 0.2 PW/cm$^2$, 0.3
PW/cm$^2$ and 0.4 PW/cm$^2$ respectively and Fig. 2 gives 0.238
a.u., 0.214 a.u. and 0.190 a.u. respectively. This implies no AC
Stark effect for the population of the two states involved in Rabi
oscillations.

Further varying the wavelength, the Rabi effect disappears rapidly.
For a 208 nm, 0.3 PW/cm$^2$ pulse, the oscillations of the
population of doubly excited states and of the flux of double
ionization are blurred severely, as shown in Fig. 4. In addition,
both the population of doubly excited states and the flux of double
ionization at 208 nm are much lower than those at 198 nm. Note that
doubly excited states can not be populated when the intermediate
resonance is broken and population in region 1 arises mostly from
contribution of doubly ionizing wavepackets. However, the
double-circle structure [see Fig. 1(c)] can be still distinguished
in the logarithmic plot of the correlated momentum spectrum, but
with probability of about two orders of magnitude lower than the
single-circle structure directly coming from the unsplit ground
state. In the corresponding total kinetic energy spectrum we find
the primary peak shifted from the middle position of the double-peak
due to the AC Stark effect of the ground state, as shown in Fig. 5.
Thereby the AC Stark shift at 208 nm, 0.3 PW/cm$^2$ is determined to
be 0.012 a.u. At shorter wavelengths (about 142 nm) or longer
wavelengths (about 228 nm), again we found the multiphoton Rabi
oscillations of the two correlated electrons.

\begin{figure}
\begin{center}
\includegraphics[width=6.5cm]{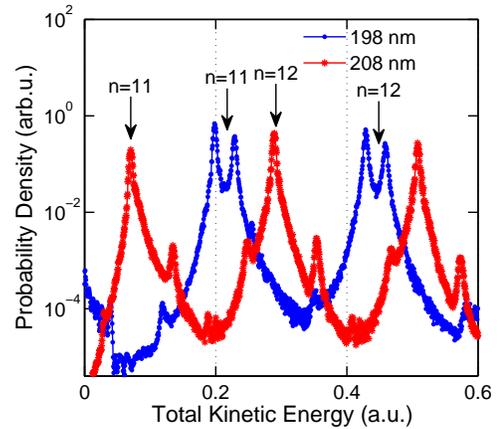}
\caption{\label{fig5}(color online) Photoelectron total-kinetic
energy spectra of two ionized electrons from double ionization of He
by 0.3 PW/cm$^2$ laser pulses at two different wavelengths.}
\end{center}
\end{figure}

In summary, we have demonstrated the multiphoton Rabi oscillations
of strongly correlated electrons in strong-field NSDI by quantum
mechanical calculations. The demonstration of the two-electron
multiphoton Rabi effect both in time domain and in frequency domain,
enables one to have a deep insight into electronic correlations and
complicated multiphoton phenomena. Our study, fundamentally
important to quantum optics and many-body physics, can advance
exploring the physical mechanisms of many effects in nature governed
by electronic correlations. The optical Rabi effect involving two
correlated electrons is expected to be a new tool for broad
applications, such as direct quantum coherent control in atomic
clock \cite{Chaudhury, Windpassinger}, quantum information
processing \cite{Li, Collin, Wallraff} and chemical reactions
\cite{Sussman}.

\begin{acknowledgments}
This work was supported by the National Natural Science Foundation
of China under Grant No. 11004070, National Science Fund for
Distinguished Young Scholars under Grant No.60925021, and the 973
Program of China under Grant No. 2011CB808103.
\end{acknowledgments}

\end{document}